\documentclass[aip,apl,preprint,twocolumn,graphicx]{revtex4-1}

\usepackage{graphicx}
\usepackage{dcolumn}
\usepackage{bm}

\usepackage{amssymb}
\usepackage{amsthm}
\usepackage{amsmath}

\draft

\begin{document}


\title{Deuterated polyethylene coatings for ultra-cold neutron applications}


\author{Th. Brenner}
\affiliation{Institut Laue-Langevin, 38042 Grenoble Cedex 9, France}
\author{P. Fierlinger}
\affiliation{Physikdepartment, Technische Universit\"at M\"unchen, D-85748 Garching, Germany}
\author{P. Geltenbort}
\affiliation{Institut Laue-Langevin, 38042 Grenoble Cedex 9, France}
\author{E. Gutsmiedl}
\affiliation{Physikdepartment, Technische Universit\"at M\"unchen, D-85748 Garching, Germany}
\author{A. Hollering}
\affiliation{Physikdepartment, Technische Universit\"at M\"unchen, D-85748 Garching, Germany}
\author{T. Lauer}
\affiliation{Forschungsneutronenquelle Heinz Maier-Leibnitz, Technische Universit\"at M\"unchen, D-85748 Garching, Germany}
\author{G. Petzoldt}
\affiliation{Physikdepartment, Technische Universit\"at M\"unchen, D-85748 Garching, Germany}
\author{D. Ruhstorfer}
\affiliation{Physikdepartment, Technische Universit\"at M\"unchen, D-85748 Garching, Germany}
\author{J. Schroffenegger}
\affiliation{Forschungsneutronenquelle Heinz Maier-Leibnitz, Technische Universit\"at M\"unchen, D-85748 Garching, Germany}
\author{K. M. Seemann}
\affiliation{Physik-Department E21 \& Heinz  Maier-Leibnitz Zentrum (MLZ), Technische Universit\"at M\"unchen, D-85747 Garching, Germany}
\author{O. Soltwedel}
\affiliation{Max-Planck-Institute for Solid State Research, Outstation at MLZ, Lichtenbergstr. 1, 85747, Garching, Germany}
\author{St. Stuiber}
\affiliation{Physikdepartment, Technische Universit\"at M\"unchen, D-85748 Garching, Germany}
\author{B. Taubenheim}
\affiliation{Physikdepartment, Technische Universit\"at M\"unchen, D-85748 Garching, Germany}
\author{D. Windmayer}
\affiliation{Physikdepartment, Technische Universit\"at M\"unchen, D-85748 Garching, Germany}
\author{T. Zechlau}
\affiliation{Forschungsneutronenquelle Heinz Maier-Leibnitz, Technische Universit\"at M\"unchen, D-85748 Garching, Germany}


\date{\today}


\begin{abstract}
We report on the fabrication and use of deuterated polyethylene (dPE) as a coating material for ultra-cold neutron (UCN) storage and transport. 
The Fermi potential has been determined to be 214~neV and the wall loss coefficient $\eta$ is 1.3$\cdot$10$^4$ per wall collision. 
The coating technique allows for a wide range of applications in this field of physics. 
In particular, flexible and quasi-massless UCN guides with slit-less shutters and seamless UCN storage volumes become possible. 
These properties enable the use in next-generation measurements of the electric dipole moment of the neutron.\end{abstract}

\maketitle





\section{\label{sec:level1}INTRODUCTION}
Ultra-cold neutrons (UCN) have velocities below $\sim$~7~m/s or few 100~neV in kinetic energy.
At these energies, their unique properties make them a useful tool in fundamental physics:
Gravity causes a change in potential energy of 1.02~$\cdot$~10$^{-7}$~eV per meter height, the magnetic moment is $\lvert\mu_n \rvert \sim$~6~$\cdot$~10$^{-8}$~eV$/$T  and the strong interaction causes repulsion from many material surfaces \cite{golub}.
Due to the large wavelength of UCN, $\lambda >$~50~nm, the wave function experiences an average coherent interaction with materials, resulting in a wave-like reflection on a potential wall with only a small probability for absorption.
This so-called Fermi pseudo potential \cite{fermi1,fermi2a} consists of a real and an imaginary part $V_F = V - iW$: 
\begin{equation}
V = \frac{2 \pi  \hbar^2}{m_n} N b                       
\end{equation}
and 
\begin{equation}
W = \frac{\hbar}{2} N \sigma v.                       
\end{equation}
Here, $m_n$ and $v$ are the neutron's mass and velocity, respectively, $N$ is the atomic number density of the wall material, $b$ is its bound coherent scattering length and $\sigma$ is the loss cross-section. 
The real part of $V_F$ determines the height of the wall potential and the ratio $\eta = W / V$ the loss coefficient, which is related to the angle-averaged loss probability per wall collision  
\begin{equation}
\mu(E_n) = 2 \eta \bigg[    \frac{V}{E_n} \arcsin \bigg( \frac{E_n}{V} \bigg)^{1/2}   - \bigg( \frac{V}{E_n} - 1 \bigg)^{1/2}    \bigg],                
\end{equation}
with the kinetic energy of the neutron $E_n$. 
For $E_n < V$, UCN are reflected from a surface under any angle of incidence.
$V$ can reach values up to 250~neV for beryllium or diamond-like carbon surfaces \cite{DLC}.
One of the flagship experiments with UCN is the search for the electric dipole moment (EDM) of the neutron \cite{ramsey}, where UCN are confined in a trap either at room temperature \cite{baker,tumedm} or at cryogenic temperatures \cite{LANL}.
The statistical sensitivity of an EDM experiment is
\begin{equation}
\sigma_d = \frac{\hbar}{2 \alpha E T \sqrt{N}},
\end{equation}
with a quality parameter $\alpha$, an applied electric field $E \sim$~20~kV$/$cm, the storage time $T \sim$~250~s and $N$ the number of detected neutrons at the end of the experiment.
Here, $V$ is important due to the typical spectrum of UCN from the source, which is typically proportional to $V^{3/2}$. A high value of $V$ therefore leads to an accordingly higher number of initially stored neutrons.
A low loss coefficient $\eta$ and low outgassing rates reduce losses during the storage period and thus lead to both longer storage times $T$ and an increased number $N$.  \\
Due to the large value of the electric field $E$, highly electrically insulating behaviour and resilience against electrostatic field breakdowns is required and again low outgassing rates are required to prevent a negative impact of residual gas on the high voltage performance.
Previously, coatings for insulating rings of UCN storages chambers were made of bulk quartz with $V = \,$ 95~neV \cite{baker, serebrov} or deuterated polystyrene \cite{psi} with $V \approx \,$ 160~neV.
In addition to possessing a relatively low $V$, these materials suffer from problems with water contamination of their surfaces or from significant outgassing in vacuum without prior baking processes. Both of these contaminations led to UCN losses due to upscattering on hydrogen or interactions with the residual gas.

Finally, to obtain a large quality parameter $\alpha$, the spin-flip probability per wall collision must be small.   Typical materials have a probability on the order of $10^{-6}$ per collision \cite{peter_depol1,peter_depol2,peter_depol3}.

One viable chamber wall material is the plastic insulator, fully deuterated polyethylene. It has a calculated scattering length density (SLD) of $8.323 \times 10^{-6}$~ \AA~ ($V = 214.7$ neV) which is an improvement over previous materials.  Our work describes the development of a technique to produce deuterated polyethylene (dPE) coatings of large surfaces and experiments which determined its relevant parameters for UCN applications.

\section{\label{sec:level2}COATING PROCEDURE}
\begin{figure}
\includegraphics[width=0.45\textwidth]{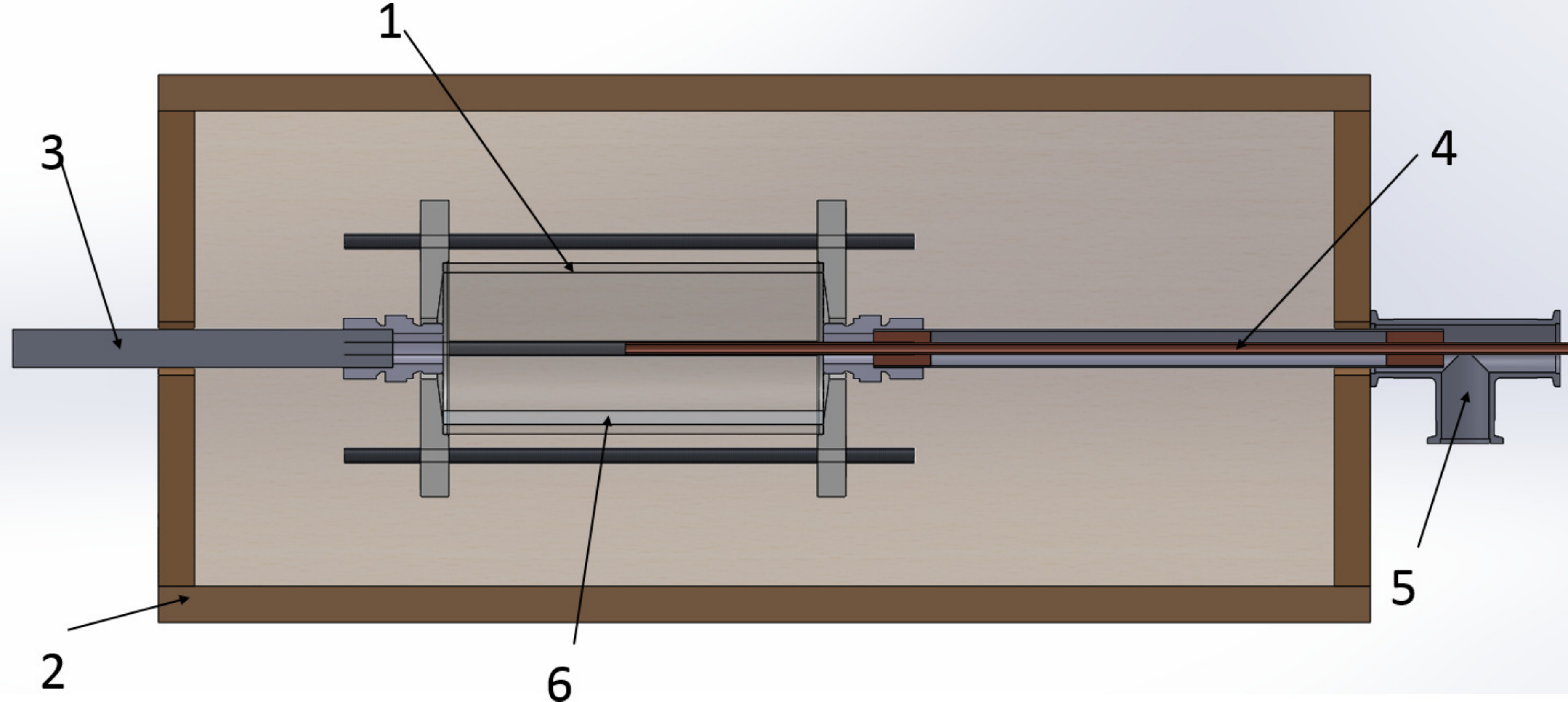}
\caption{\label{fig:fig1} Schematic view of the coating setup. The sample [1] was placed inside an oven [2], which was heated with air. The sample was connected to a stepping motor by the axle [3]. The tube was flushed with hot N$_2$ through the inlet [4] and the evaporated solvent was recovered by recondensation [5]. For illustration purposes, the dPE/xylene ``puddle" is also shown [6].}
\end{figure}

The general idea of the coating procedure is to dissolve dPE powder in a deuterated solvent (d-xylene), wet the surface to be coated with the solution, and let the solvent evaporate\cite{kuzniak}. 
Polyethylene in general only dissolves at temperatures around 145$^\circ$~C, which is also very close to its melting point. This requires that the procedure must take place inside an oven. 
A schematic view of the coating apparatus for cylindrical glass tubes is shown in Fig.~\ref{fig:fig1}. 
It allowed us to coat the walls of glass tubes with diameters of 80-125~mm and lengths of up to 500~mm.
To remove any dirt on the tube surface, they were cleaned thoroughly by hand with isopropanol and acetone prior to the coating procedure. 
Afterwards, the tubes were closed on both ends with Teflon caps, since Teflon is chemically inert when exposed to xylene. 
The tubes were then baked at 100$^\circ$~C for about an hour to remove any remnants of the cleaning agents,flushing them with N$_2$ at the end.
Two grams of deuterated polyethylene powder (Polyethylen-d4 98 Atom$\%$D, SigmaAldrich) and 500 ml deuterated xylene (xylene-d10, ARMAR Chemicals) were filled into the cylinder through a 1" hole in one of the end caps.
The entire assembly was then mounted inside the oven in such a way that it was freely rotatable about its cylindrical axis. 
On one side, it was connected to the axle of a stepping motor, on the other side there was an opening that allowed us to flush the setup with hot (approximately 160$^{\circ}$~C) N$_2$ during the coating procedure. The hot gas could flow into the setup through a copper tube with diameter 6~mm and stream back out through a surrounding steel tube with diameter 20~mm. The oven was closed and heated to approximately 145$^\circ$~C. 
After the dPE powder dissolved completely, the tube rotated with a speed of a few rotations per minute. 
Typically, a few large scale structures of ``melted" dPE appeared on the walls but subsequently disappeared as they repeatedly passed through the solvent ``puddle" on the bottom of the tube. 
The rotation continued until all large scale structures disappeared and it was apparent that the wall was wetted homogeneously with the solution. 
The volume was then flushed with nitrogen to remove the evaporated xylene from the system. 
The xylene re-condensed in a tube and was recovered for later use. 
Once all of the solvent was gone, the heating was turned off and the system cooled down. 
After the system reached room temperature, the tube was removed from the system.
In addition to these tubes, rings with diameter 500~mm and height 120~mm, typical for a cylindrical EDM storage chamber, were also coated with this procedure.

Depending on the substrate and the coating thickness, the dPE coatings showed quite different behaviour. Although all coatings tended to stick well to the surfaces, it was possible to remove whole coatings from smooth glass tube substrates without damage.
For rough (optically opaque) quartz substrates and thinner ( $<$~5~$\mu$m) coatings, removal of the coatings was only possible using the inverse coating procedure with hot solvents.
\section{\label{sec:level3}SAMPLE CHARACTERIZATION}

\subsection{\label{subsec:level3}Outgassing, UV transparency and magnetic properties}
As mentioned in section I., low outgassing of the UCN storage chamber is important since it is not actively pumped during the storage time. 
Coated tubes with diameters of 80 and 115 mm were placed without endcaps into a 0.5~m$^3$ vacuum chamber and pumped down into the 10$^{-7}$~mbar range.
We obtained rates for both tubes on the order of 10$^{-8}$~mbar l s$^{-1}$ cm$^{-2}$ despite the different surface areas. 
This indicated that the outgassing was dominated by the vacuum chamber and the rate due to dPE-coated walls was less than 10$^{-8}$~mbar l s$^{-1}$ cm$^{-2}$, which is sufficiently low for UCN storage experiments.

If dPE is to be used as wall coating material in an nEDM experiment operating with a mercury co-magnetometer, as described in Ref.~\onlinecite{baker}, it is important that it is not opaque to ultra-violet light. We measured the transparency of a 10~$\mu$m thick foil with a 254 nm laser and determined it to be roughly 60\% transparent. Optical windows coated with a thin layer of dPE (e.g. via spin-coating\cite{kuzniak}) therefore would not significantly impact the performance of the Hg co-magnetometer.

Since nEDM measurements are performed with polarized UCN, the wall coating material must be non-magnetic. To test this, we measured a dPE sample with a SQUID apparatus\cite{squid} at 10 cm distance in the magnetically shielded room \cite{shielded_room} at a sensitivity of 10 pT. No signal above background was observed.

\subsection{\label{subsec:level3a}Surface roughness}
To obtain a characteristic dPE sample, the coating was removed from one tube by first pulling it carefully off the edges of the tube. The rest of the coating was then peeled off the glass, resulting in a 10~$\mu$m thick foil.
An atomic force microscope measurement was performed on both sides of the foil. It showed roughnesses of several 100 nm on the ``inner" side (exposed to air during the coating) and about 10 nm on the ``outer" side (attached to the glass wall). It should be noted that this unevenness occurred on a $\mu$m scale (due to the dPE forming ``droplets"), which is important for the reflectivity measurements discussed in the next section. The inset in Fig.~\ref{fig:fig5} shows a height profile of one such measurement.
The measured values are both suitable for storage experiments and can be improved significantly by applying a more thorough cleaning technique, such as has been used to prepare NiMo coated glass tubes \cite{nimo_coatings}. 
Additionally, moving the oven into a clean-room environment will reduce the amount of dust particles and improve surface roughness on the outer side.

\subsection{\label{subsec:level3b}Fermi potential}
\begin{figure}
\includegraphics[width=0.45\textwidth]{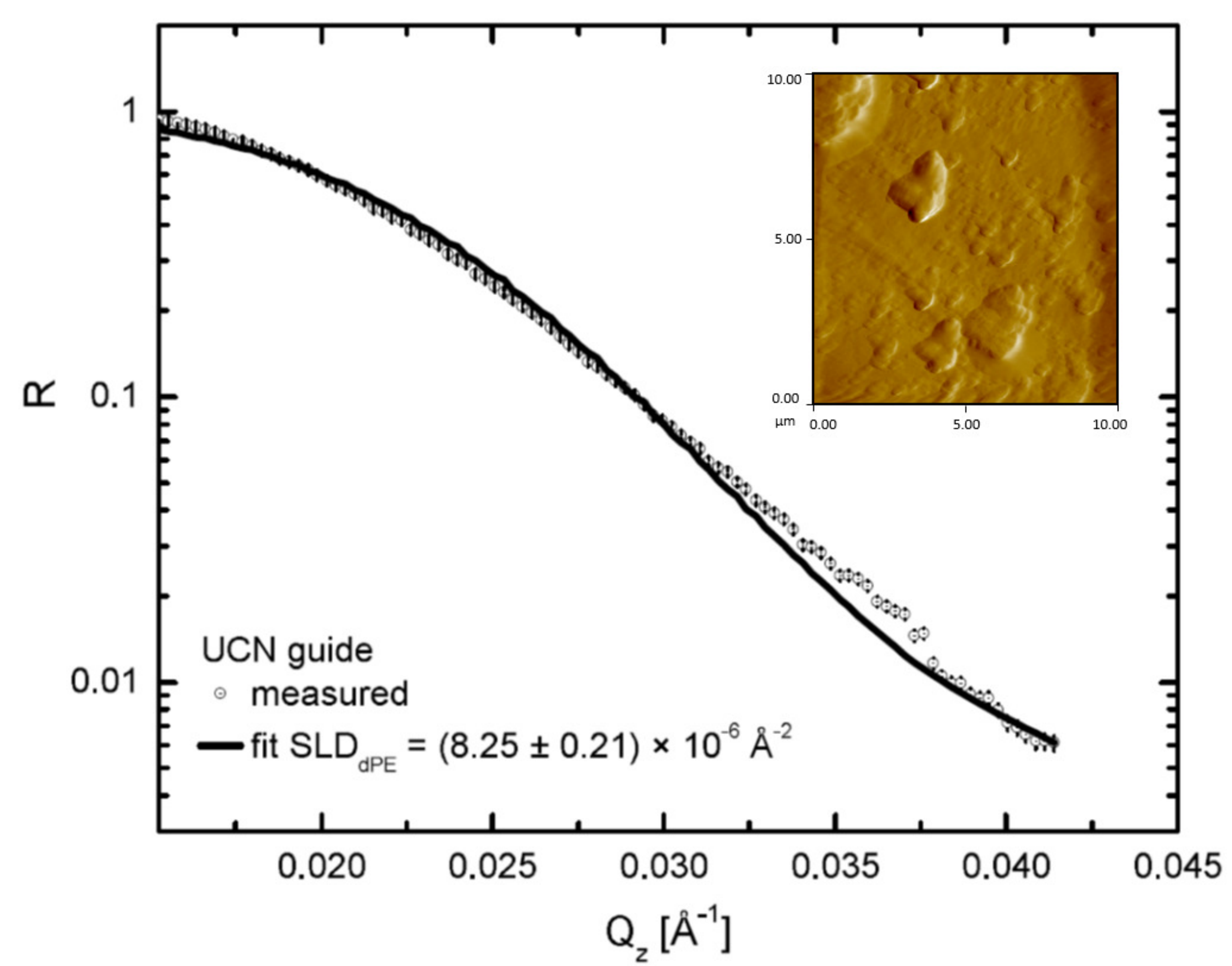}
\caption{\label{fig:fig5} Reflectivity vs. momentum transfer of cold neutrons scattered off dPE coated guides. The solid line is the result a very simple two-layer fit, the uncertainty is the one returned by the fit routine. The inset shows a picture of an AFM measurement. The peak-to-peak height on that particular measurement is several 100 nm. 
}
\end{figure}

The critical angle of the wall coating was measured by cold-neutron reflectometry using the MIRA instrument at the Forschungs-Neutronenquelle Heinz Maier-Leibnitz (FRM-II) research reactor in Garching. The cold-neutron beam (wavelength 4~\AA) was scattered directly off the inner surface of the coated tubes used for UCN storage measurements. Systematic effects due to the curved nature of the guides were minimized by collimating the incoming neutron beam to an area of 1$\times$1~mm$^2$.
A reflectivity curve is shown in Fig.~\ref{fig:fig5}.
Neutrons reflect specularly from flat surfaces, but as the dPE-surface exhibited $\mu$m scale raised regions (see previous section), the reflected beam was partially divergent. Thus, depending on the real incident angle of each neutron and the Fermi-potential gradient at the interface, some neutrons were reflected totally while others were not. The result is a smearing of the critical edge, as observed by neutron reflectivity. We modelled this effect by treating the data as if an incoming divergent beam was reflected on a sharp and flat interface, since this case is mathematically treated in the same way. Therefore we modelled the reflectivity curve with a dramatically increased angular resolution of 0.45$^{\circ}$ at a sharp and flat dPE/air interface.
With this method, we obtained a value for the SLD of (8.25$\pm$0.21)$\cdot$10$^{-6}$~\AA$^{-2}$, resulting in a V$_F$ of (214.8$\pm$5.2)~neV.
The uncertainties are those returned from the fit procedure and do not take into account additional systematic effects such as curvature of the guides or surface contaminants and are shown for the sake of consistency. The determined value agrees very well with the $V$ deduced by the UCN density at $t = 0$ in storage experiments compared with NiMo 85-15 (see next section) and with  $V$ determined with flat samples measured at PSI\cite{psi} of 214$\pm$10~neV.

\subsection{\label{subsec:level3c}UCN storage measurements}
\begin{figure}
\includegraphics[width=0.45\textwidth]{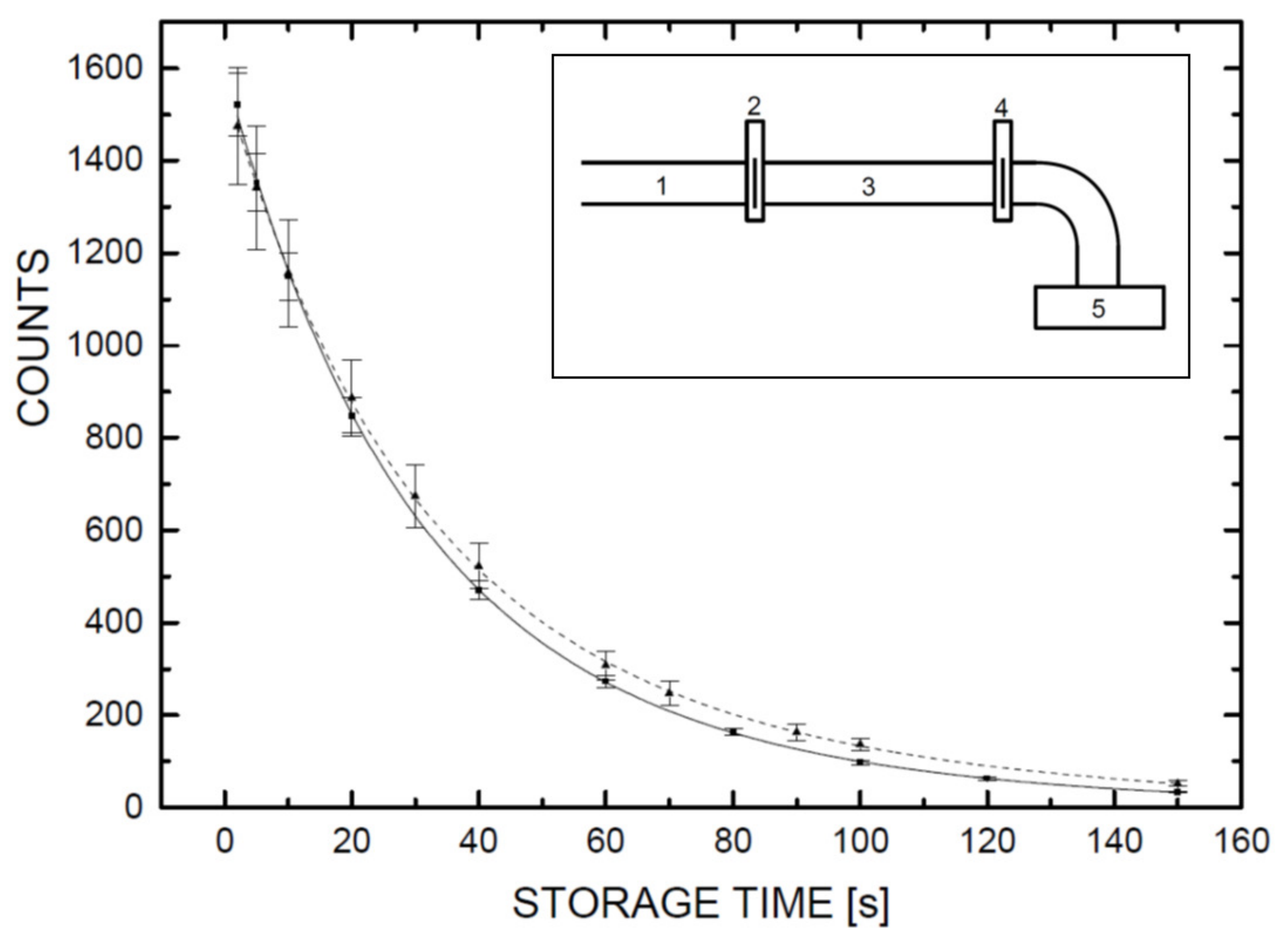}
\caption{\label{fig:fig3}Number of trapped UCN as function of storage time for tubes of identical size coated with NiMo 85-15 (solid line) and dPE (dashed line). The trap was comprised of the coated tubes and shutters coated with NiMo 85-15 at the ends. The measured storage times are ($52\pm 4)\,$s and $(61 \pm 6)\,$s for NiMo and dPE, respectively. 
Inset: Schematic view of the setup to measure the loss coefficient of the dPE coating. Neutrons entered the setup from the UCN turbine at ILL-PF2 [1]. The shutters [2] and [4] could be closed to trap the UCN inside the coated glass guide [3]. The UCN were emptied at the end of the storage cycle into the detector [5].
}
\end{figure}

A coated tube with inner diameter = 115~mm and length = 500~mm has been tested using trapped UCN at the PF2 instrument at ILL \cite{PF2}.
For these measurements, the tube was closed on both sides with NiMo coated end caps. UCN were filled into a storage section, which could be closed on both ends. After a given time period, the UCN were emptied into a UCN detector and counted. A schematic of the setup and the resulting UCN storage curve is shown in Fig.~\ref{fig:fig3}. 
The cleaning of the coating was done by manually wiping the surface. No baking in vacuum or other treatment was applied. 
The vacuum during the measurements was in the 10$^{-6}$~mbar range.
It should be noted that due to the non-polarity of dPE, the layers are hydrophobic.

At $t$ = 0, the interpolated density of stored UCN was about 0.3~cm$^{-3}$, which is comparable to results obtained with a fully NiMo coated volume. Since they were tested in the same reactor cycle, the similar neutron densities implied that $V_{F,\text{NiMo}} \sim V_{F,\text{dPE}}$.
The spectrum of the UCN coming from the source included neutrons with kinetic energies which are greater than $V_{F,\text{dPE}}$ and were thus only marginally trapped and were lost over the first $\sim$~30~s of the storage time.
To account for this ``cleaning" time, the data was fitted with a sum of two exponentials, which is a sufficient approximation for this purpose:
\begin{equation}
N(t) = A_1e^{-t/\tau_1}+A_2e^{-t/\tau_2}
\end{equation}
Here, the number of neutrons $N(t)$ is dominated by the storage time constant $\tau_1$ for the marginally trapped neutrons and $\tau_2$ for the UCN with $E < V$ and we obtain $\tau_2$~=~61$\pm$6~s from the fit.
The surface ratios inside the volume are 180641~mm$^2$ dPE coated surface, 20773~mm$^2$ NiMo 85-15 coated end caps (0.1 fraction of the total surface) and 20~mm$^2$ (7$\cdot$10$^{-5}$ of the total surface) fully absorbing slits at the connections of the end-caps.
Therefore, the storage time is composed of
\begin{equation}
\frac{1}{\tau_{\text{meas}}} = \frac{1}{\tau_{n}} + \frac{1}{\tau_{\text{slit}}} + \frac{1}{\tau_{\text{NiMo}}} + \frac{1}{\tau_{\text{dPE}}}
\end{equation}
$\tau_n$ causes 7$\%$ of the neutrons to decay within the bottle during 60 s.
The storage life-times caused by the material surfaces are $\tau_{\text{mat.}} = 1 / \left(\mu_{\text{mat.}}(E) \nu\right)$, where $\nu$ is the wall collision frequency and $\mu(E)$ the energy dependent loss per wall collision.
With $V$ = 214~neV (see Sec.~\ref{subsec:level3b}), we obtain $E_{avg}$ = 142~neV and from kinetic gas theory we calculate $\nu =  v /\lambda  = 46$~Hz where $\lambda$ = 115~mm is the mean free path of the UCN and the velocity $v$ = 5.2~m$/$s. These values account for the different collision frequency on the NiMo coated end-caps and the dPE-coated walls of the tube.
The loss coefficient $\eta$ is $\sim \mu / 1.5$ for our spectrum and the assumed $V$. 
To estimate the loss on the end-caps, we conservatively determine $\mu(E)_{\text{NiMo}}$ = (2.4$\pm$0.3)$\cdot$10$^{-4}$ from the measured storage time in NiMo coated tubes.
Then, we obtain the value for $\eta_{\text{dPE}}$ = (1.3$\pm$0.3)$\cdot$10$^{-4}$.

\section{\label{sec:level4}DISCUSSION AND CONCLUSION}
We have demonstrated that UCN storage can be performed by coating large surfaces with dPE. 
The material is superior to other deuterated plastic coatings (e.g. dPS), due to its high Fermi potential, high transparency for UV light, low outgassing, high mechanical strength, high resistivity, resilience to electrostatic breakdowns, hydrophobic behaviour and simple usability. 
In previous approaches, only comparably small samples of dPE coated surfaces were realized such as using the spin-coating technique for window coatings in an EDM experiment\cite{baker}. The coating technique presented here is comparably simple, versatile, safe and reliable and able to provide coatings for large surfaces.
A variety of applications can be envisaged:
\begin{itemize}
\item{Coating of UCN storage volumes without slits at mechanical joints:}
By rotating the storage volume during the coating process in more than one direction or by thoughtful geometric arrangements and using a sequential coating procedure, it is possible to produce volumes with all mechanical joints and slits coated and filled with dPE. Scaling from the measured loss coefficient, UCN storage life-times inside a typical EDM chamber of 125~s to 250~s can be obtained if the experiment is positioned at the PF2 instrument \cite{PF2} or a new source\cite{oliver}, respectively. Both values can be achieved without any preconditioning of the surfaces in vacuum.
\item{Flexible guides:}
Since the coatings can be removed from a smooth surface if produced with larger than about 5~$\mu$m thickness, guides with flexible sections can be realized. 
An application for such a configuration is a guide with an integrated shutter without any joints of components and thus no slits.
\item{dPE replica tubing:}
After removing the coatings from the smooth surface, the smooth side can be inverted to obtain a smooth inside of the volume (e.g. a seamless tube).
\item{Radiation-hard quasi-massless guides:}
dPE is radiation hard and, as a foil, has a very low mass. This makes it ideal for use in a UCN guide to transport UCN from an in-pile UCN source to outside of the biological shield.
\end{itemize}
We wish to acknowledge the support at the Maier-Leibnitz Zentrum (MLZ) in Garching and at the Institute Laue-Langevin in Grenoble. This work was supported by the DFG Cluster of Excellence ``Origin and Structure of the Universe" and the DFG Schwerpunktprogramm 1491 ``Precision Experiments in particle- and astrophysics with cold and ultracold neutrons". The authors also thank A. Mantwill and R. Schwikowski for technical assistance.

\bibliography{apssamp}
\bibliographystyle{model1-num-names}

\end{document}